# Validity-first automatic polycube labeling for CAD models

Sébastien Mestrallet[*†]     Christophe Bourcier[*]     Franck Ledoux[†‡]

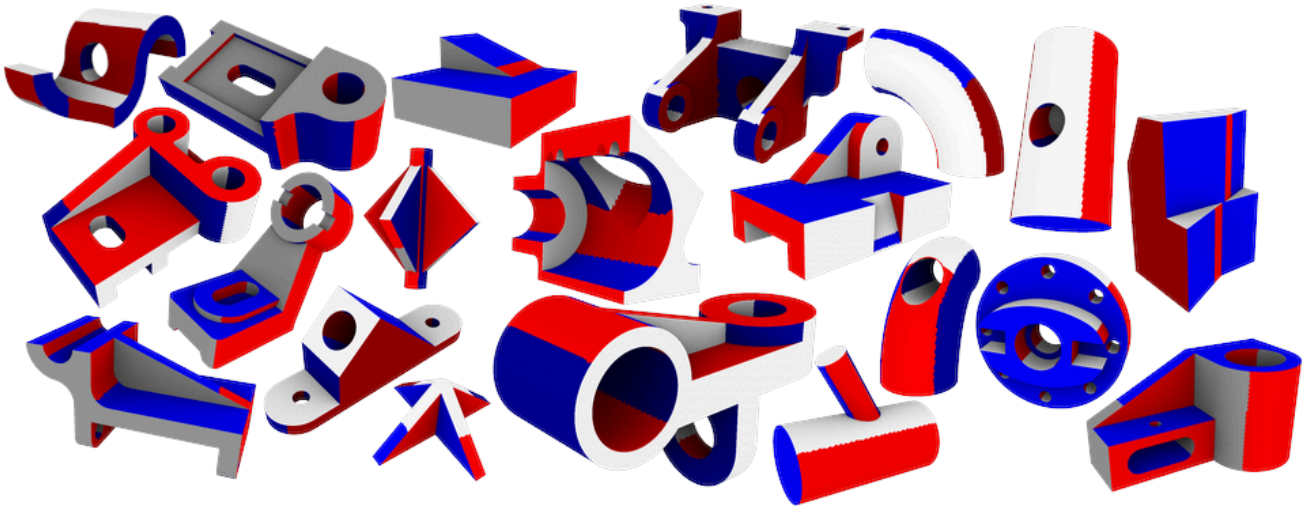

Figure 1: Collection of polycube labelings generated using our algorithm.

**Abstract**

For many simulation codes, block-structured hex meshes remain preferred while their automatic generation is unsolved. We investigate the usage of a polycube-based approach. More specifically, we focus on the labeling stage, which consists in assigning each boundary facet to one of the 6 signed principal axis. Similar works are confronted with 2 challenges: overconstraining validity criteria, and the conflated processing of validity criteria with quality metrics. We tackle these obstacles with automatic routines based on semi-global labeling operators. Our approach is successfully tested on CAD models, which are of interest for many numerical simulation problems.

## 1 Introduction

Hex meshes, and more specifically block-structured hex meshes, are preferred for high-fidelity numerical simulation in several fields like computational fluid dynamics (CFD), hypersonic flows, hydrodynamics, or structural analysis [1]. They are efficient in highly anisotropic physical simulations (boundary layers, shockwaves, etc.), as the associated tri-linear basis has cubic terms that capture higher order variations and provide less elements, reducing simulation time. Such types of simulations frequently consider geometric domains that are CAD (Computer-Aided Design) models made of points, curves, surfaces, and volumes with sharp features. For complex assemblies of mechanical parts, state-of-the-art hex meshing algorithms [2] are not yet applicable. As a consequence, industry-mature tools produce hex block structures by adopting an interactive decomposition strategy [3], [4]. Two main strategies are possible: either the CAD shape is split at the geometric level into simple shapes that can be discretized using sweeping-like algorithms (Abaqus, Cubit/CSimsoft); or the CAD shape remains unchanged, and an associated structure is split into hex blocks (ICEM-CFD Hexa). Both options require interactive tools with sophisticated GUI, as well as qualified engineers that spend hours or days to create meshes. Today, this process is a bottleneck for incremental design and sensitivity analysis.

Hex meshing algorithms have been struggling for several decades regarding the global combinatorial structure of hex meshes that prevents algorithms to rely on local decisions [5]. To provide an automatic solution, research on hex meshing focuses on global approaches where the block structure may be extracted directly using frame fields, or indirectly by generating a non-optimal structure, using polycubes or medial axes, that are optimized thereafter. Grid-based [6] and octree-based methods [7], [8] may also be used as an indirect approach but their lack of regular structure makes difficult the topology optimization post-process. Medial structures have been used to guide all-hex meshing for simple CAD models [9], tubular-like shapes [10], [11] or more general CAD models using post-process operations and sweeping [12]. The difficulty of robustly generating a clean medial object for 3D CAD models, coupled with the necessity to simplify it, makes this approach still challenging. The idea of polycube-based approaches is to "deform" a volumetric mesh into a polycube shape [13], [14] by defining a polycube-map from which integer grid lines will be used to extract regular grids. Such approaches have been demonstrated as being efficient and automatic. However, due to the lack of inner singularities, obtained meshes may suffer from large distortions near the boundary, and sharp features of CAD models are difficult to handle. Unlike polycube, frame-field-based approaches aim to first insert singularities inside the domain in order to build global parameterizations with low distortion along the boundary. Such parameterizations can be obtained from a frame field using *CubeCover* [15] or *PGP3D* [5], [16]. The main issue is that an arbitrary 3D frame field can have a topological structure that is not compatible with hex meshing. And unfortunately, generating hex-compatible frame fields is still challenging.

[*]Université Paris-Saclay, CEA, Service de Génie Logiciel pour la Simulation, 91191 Gif-sur-Yvette, France
[†]Université Paris-Saclay, CEA, Laboratoire en Informatique Haute Performance pour le Calcul et la simulation, 91680, Bruyères le Chatel, France
[‡]CEA, DAM, DIF, F91297 Arpajon, France

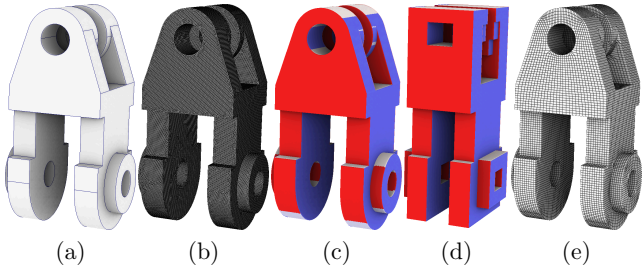

Figure 2: Our pipeline steps: (a) CAD model; (b) tetrahedral mesh $T_\Omega$; (c) polycube labeling $\ell$ on the surface mesh $\partial T_\Omega$; (d) polycube obtained by morphing the triangle mesh with the labeling orientation constraints; (e) hex mesh $H_\Omega$ obtained by quantization and inverse deformation.

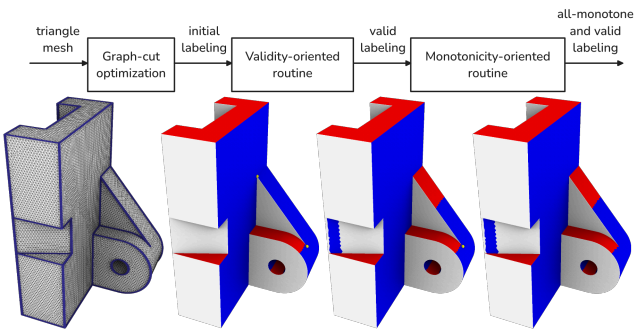

Figure 3: Diagram of our method. After a graph-cut optimization on the input mesh to generate an initial labeling, two routines are executed. The first one aiming at a valid labeling, the second at removing turning-points (in yellow) while keeping a valid labeling. S36 model from *MAMBO* [17].

We propose a new polycube algorithm dedicated to CAD models. Our approach is based on the traditional pipeline (Fig.2), which consists in: (1) discretizing the CAD model $\Omega$ with a tetrahedral mesh $T_\Omega$; (2) extracting the boundary surface $\partial T_\Omega$; (3) computing a valid and feature-preserving labeling $\ell$, which assigns boundary triangles to one of $\{\pm X, \pm Y, \pm Z\}$; (4) deforming the mesh with respect to the orientation constraints; (5) quantizing the deformed mesh within an integer grid, obtaining a polycube; (6) applying of the inverse deformation. We focus in this paper on the labeling stage, which remains very challenging for realistic CAD models. We propose an automatic procedure that relies on : improved validity criteria (Section 4); a graph-cut based initial labeling (Section 5); a routine targeting a valid labeling (Section 7.1); another one targeting an all-monotone labeling (Section 7.2); both based on existing and new semi-global operations (Section 6). In this way, our method makes a clear distinction between desirable and required conditions. Fig.3 illustrates the three steps of this algorithm. Our approach takes into account CAD feature curves as constraints to preserve. For all the other stages, we rely on established state-of-the-art components: tetrahedrization with *Gmsh* [18] or *MeshGems* [19], both available in SALOME [20]; hex-mesh extraction with *HexEx* [21] or Protais *et al.* 2022 [22]. We have tested our approach on a wide collection of CAD models, and we obtain valid and expected labeling for realistic CAD models and some configurations of features edges known as difficult to handle in an automatic manner. Our implementation is open-source and available at https://github.com/LIHPC-Computational-Geometry/validity-first-polycube-labeling.

## 2 Related works

**Polycube generation.** Polycubes were introduced in computer graphics in [14] to generate seamless texturing of triangulated surfaces. Polycubes rely on two main concepts: an axis-aligned polyhedral structure and a volumetric map $M$, often called *polycube-map*. Considering a geometric domain $\Omega$, and a tetrahedral mesh $T_\Omega$ of $\Omega$, the first stage of polycube-based approaches consists in *labeling* each triangle of $\partial T_\Omega$, which is the triangular mesh of the boundary, with a value that represents one of the six principal axes $\{\pm X, \pm Y, \pm Z\}$. A naive labeling can be computed by assigning to each surface triangle the label closest to its normal [13]. However, this does not produce a valid structure in general and labeling improvement is necessary. This improvement aims to ensure a set of sufficient topological conditions [23].

First usages of polycubes for hex meshing [13], [24] built the inverse of $M$ by applying a volumetric deformation [25], [26] to convert $T_\Omega$ to a polycube while minimizing the distortion introduced by the deformation. Then, many works have focused on improving the mesh quality by trying to reduce the number of polycube facets and the distortion of $M$, such as: reformulating the polycube building problem as a graph-cut optimization [27]; minimizing the $\ell_1$ norm of mesh normals, using an as-rigid-as-possible deformation [28]; driving the polycube to be aligned along a prescribed frame field [29]; using a constrained voxelization to generate the polycube surface [30]; using evolutionary algorithms, to explore a larger space of polycube structures [31], [32]; or by considering CAD models by adopting a two-stage process, where sharp features are first considered then the whole domain, by resolving two non-linear optimization problems [33].

**Polycube improvement.** As singularities are located on $\partial T_\Omega$ and their number is difficult to control, obtained hex meshes may suffer from large distortions near the boundary. Possible improvements are to insert new layers of hexahedra and so singularity points inside the domain using global [13] or selective padding [34]. Such processes provide more degrees of freedom to move mesh vertices while pushing singularities inside. Authors of [35] define three types of fundamental layers that can be added locally to better capture sharp boundary curves and surfaces solving an integer linear program. In a similar manner, [33] enhances polycube-maps using clever cuts directly on the polycube to capture sharp features. Other approaches rely on the base complex structure in order to post-process hex meshes in order to extract coarse structure. Such algorithms can be applied to polycube resulting meshes.

**Validity criteria.** Several research works use the topological properties of simple orthogonal polyhedra [23] to check the validity of a potential polycube. However, this set of properties is neither sufficient nor necessary as shown in [36] or [37].

**CAD sharp features.** Unlike smooth surfaces, CAD models contain sharp features (corner, ridges) where $G^1$-continuity is not guaranteed. It is a constraint to ensure when trying to get a valid high-fidelity polycube. To our knowledge, only authors of [28] and [33] explicitly consider sharp features in order to constraint the polycube structure to integrate them.

## 3 Problem statement

In this section we focus on the labeling stage of the pipeline (Fig.2), and the indicators to take into account to product valid and low-distortion labelings.

### 3.1 Polycube labeling definition

Instead of manipulating the volume to define a polycube transformation, labelings offer a simpler approach by only considering the surface. Each triangle is mapped to one of the six signed principal axes $\{\pm X, \pm Y, \pm Z\}$. The represented polycube is the one where all triangles are warped so that their normal coincide with their assigned label. In subsequent illustrations, $X$, $Y$ and $Z$ axes are respectively colored in red, white and blue, with a darker tone for the negative direction. Unless stated otherwise, figures are of our own work and rendered with *Geogram* [38].

### 3.2 Labeling graph

From the triangle-level representation, we assemble a graph by aggregating adjacent triangles having the same label as a *chart*, named $\mathcal{C}_i$. For that, we use a disjoint-set (union-find) algorithm. For two adjacent charts, the delimitation is an ordered set of edges called *boundary* ($\mathcal{B}_i$). Vertices where several boundaries meet are called *corners* ($\mathcal{V}_i$). To assemble boundaries and corners, we iterate over all mesh vertices until finding a corner (three or more adjacent labels), then we move along adjacent boundaries with a depth-first exploration.

Charts, boundaries and corners are sometimes called patches, arcs and nodes [39]. Labeling graphs are sometimes called patch layouts, or polycube layouts if valid [40]. To mention a chart, a boundary or a corner without specifying the kind of element, we later use the term *labeling graph component*. The *valence of a chart* refers to the number of adjacent charts, and the *valence of a corner* is the number of adjacent boundaries, which by definition is greater than 2.

### 3.3 Validity criteria

Polycube labelings being less restrictive than polycube deformations, work has been conducted to define necessary and sufficient validity criteria. Authors of [23] extend Steinitz's theorems [41] to orthogonal polyhedra by analyzing the graph of those polyhedra. Their analysis applies to unsigned axes mapping. Each chart is either matched to $X$, $Y$, or $Z$. Among the special kinds of polyhedra they defined, *simple orthogonal polyhedra* are "*three-dimensional polyhedra with the topology of a sphere in which three mutually-perpendicular edges meet at each vertex*". Criteria on labeling graph components (Section 3.2) can be derived as: (1) all charts must have a valence of at least 4; (2) all boundaries must be orthogonal separations, i.e. must be between charts of different axes; (3) all corners must have a valence of 3 (Fig.4).

Despite not crafted for polycubes, these criteria are popular for labeling-based polycube generation [37], for their simple expression and because the last two criteria are purely local. There are proved to be sufficient for genus-0 shapes (with unsigned labels), but are not necessary (Fig.5 left). In case of signed labelings, authors of [36] showed they are not sufficient for genus-0 neither (Fig.5 right). Those issues are discussed in *Evocube* [31] appendix A and [37]. The latter and [32] also

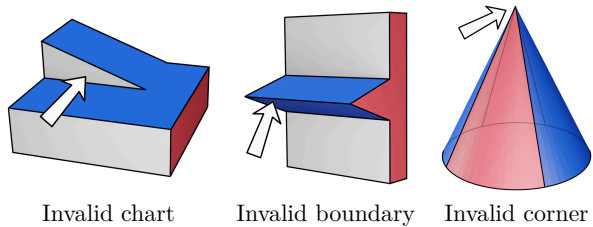

Invalid chart    Invalid boundary    Invalid corner

Figure 4: Labelings that do not correspond to simple orthogonal polyhedra [23]. Illustration from [31].

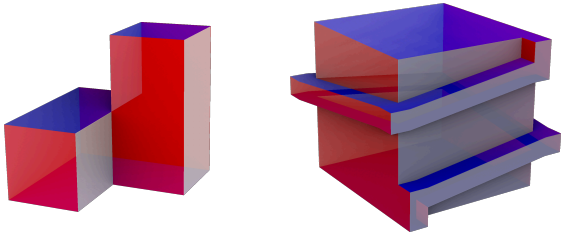

Figure 5: (left) 4 cubes joined to form a valid polycube with a 4-connected corner ; (right) A labeled genus-0 shape satisfying criteria from [23] despite not leading to a polycube if the labels' sign is enforced: cuboid slope branches will be mixed-up on the same z-coordinates.

picked another counterexample up, regarding non-orthogonal boundaries: in case of a local parametrization, overlapping polycube faces can produce distinct hex mesh surfaces and are valid when the solid angle is greater than 180° (Fig.6).

For want of anything better, *simple orthogonal polyhedra* [23] have often been used as topology constraints for polycube generation. This leads recent works to propose a specific set of conditions tailored to polycubes. *Zhao et al.* [39] focused on higher genus shapes and non-simply connected faces, but they maintain the valence-of-3 limit on corners. By counting the number of corners $V_i$ with a quad-mesh valence of i, and the number of chart corners $T_j$ having a polycube angle of $j\frac{\pi}{2}$, *He et al.* [32] expand the valid space of polyhedra for polycube generation, allowing configurations in Fig.5 left and Fig.6 right. The corner validity criterion is here constrained by a maximum valence of 6.

### 3.4 Quality metrics

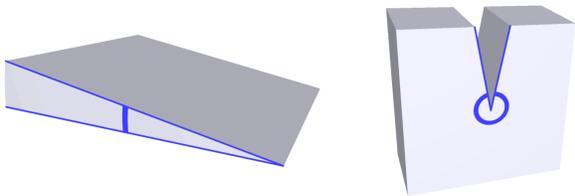

(a) solid angle < 180°  (b) solid angle > 180°

Figure 6: Two shapes with non-orthogonal boundaries: (a) rightfully invalid, the corresponding polycube has no volume, so no mesh; (b) the corresponding polycube has overlapping faces but this configuration is valid. Illustration from [37].

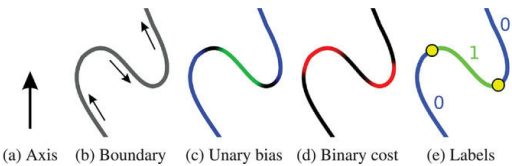

(a) Axis  (b) Boundary  (c) Unary bias  (d) Binary cost  (e) Labels

Figure 7: Identification steps for turning-points. Illustration from [31].

Because the labeling is upstream to the polycube, to avoid the costly polycube generation, we need quality metrics that are defined on the labeling. As a proxy to the geometric closeness, the per-triangle *fidelity* measures the angle between its normal and its assigned direction. The global fidelity averages the per-triangle fidelity on the surface. To favor simple polycubes and as a result low singularity count on hex-meshes, we measure the number of labeling corners $\mathcal{V}_i$, and to favor straight boundaries – avoiding parametric distortion along them – we identify vertices called *turning-points* where the boundary markedly drift away from the assigned direction. In the same manner as [27], [31], we use a graph-cut optimization with two possible labels per edge (Fig.7). A turning-points-free boundary is said to be *monotone*. An all-monotone labeling is a labeling having no turning-points. To deal with CAD models, our algorithm takes into account feature edges; a low distortion polycube should preserve them as much as possible, by coinciding feature edges and labeling boundaries. This goal can be measured by counting how many feature edges have different labels on either side. Such feature edges will be said "*preserved*" whereas those having the same label on both sides will be said "*lost*" (Fig.8).

## 4 Improved validity criteria

Our algorithm proposes the optional acceptance of boundaries between opposite labels, but along > 180° solid edges (equivalent to boundaries between 3-connected, $V_6$ and $T_4/T_1$ corners from [32]). Our proposed criterion on boundaries is therefore: "*A boundary $\mathcal{B}_i$ must be between charts assigned to different axes. Optional: or between charts of the same axis if the solid angle is greater than 180°*".

In the case of corners validity, [37] showed that high-valence corners (greater than 3) can be valid, and these configurations arise in academic datasets. Recalling the axis assignment (among $\{X, Y, Z\}$) on polycube boundaries, we can see the difference between truly invalid corners and valid ones in Fig.9. On the top left, we have 4 Z boundaries meeting at the corner. In the polycube domain, 4 parallel lines to the Z axis cannot meet at a vertex. On the top right, we have 2 X and 2 Z axes, so for each axis we have a pair of boundaries where one of them is an elongation of the other one. This analysis can be extended to higher valence corners, for example on the bottom row with 6-connected corners, where the boundaries can be grouped in pairs on the right, but not on the left.

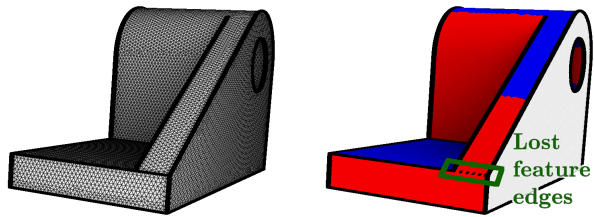

Figure 8: (left) a triangle mesh with feature edges (thick lines); (right) a labeling with most feature edges preserved (on boundaries), except inside the green box (same red label on both sides). Model S35 from *MAMBO* [17].

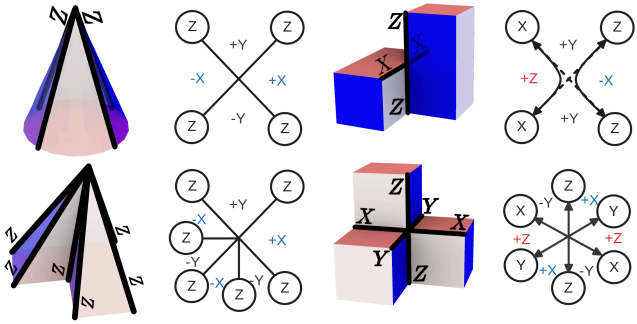

Figure 9: Better discrimination between valid and invalid corners: (top) 4-connected corners (bottom) 6-connected corners (leftmost two columns) invalid corners (rightmost two columns) valid corners. For each shape, the diagram on the right depicts adjacent labels and boundaries.

Our corner criterion no longer constrains its valence, but the associativity of incident boundaries. Boundary axes must not be the same, else we allow 2 pairs of Z axes, like in the cone case Fig.9 top left. However, we must not forget the trivial case of a cube corner, and accept trio of $XYZ$ boundaries. Our proposed formulation is "*The incident boundaries of a corner $\mathcal{V}_i$ must not be associated to the same axes, they must be associable in pairs, or make a $XYZ$ trio*". This boundaries processing at corners allows us to reach the same polycube feasible space as [32] – accepting higher connectivity, higher singularity count ($V_i$), and $T_2$ and $T_4$ points (following their convention). Nevertheless, our approach is not limited to 6-connected corners (e.g. [37, fig.5 & 8d]).

## 5 Initial labeling

The most straightforward polycube labeling generation consists in picking, for each triangle, the nearest label $\in \{\pm X, \pm Y, \pm Z\}$ of its normal. We refer to this method as the *naive labeling*. The obtained labeling is not guaranteed to be valid, the compactness can be too low or too high depending on the geometry, but the fidelity is maximal. An important issue with this initial labeling is the fragmentation induced on sub-surfaces poorly aligned with principal axes (Fig.10a). Indeed, the closest label assignment is subject to floating-point numbers inaccuracy, and the uniqueness of the result: if the normal at $\partial\Omega$ is locally $\vec{n} = \left(\frac{1}{2}, \frac{1}{2}, 0\right)$, the closest label on $\partial T_\Omega$ can be just as $+X$ as $+Y$. There is no reason to prefer one over the other, except to impose global consistency. With the naive labeling however, the preference only derives from the floating-point representation of coordinates.

To avoid labeling fragmentation, *PolyCut* [27] and *Evocube* [31] feed boundary triangles and their normals into an optimization process aiming at higher compactness by minimizing cuts [42], [43], [44], the so-called *graph-cut optimization* (Fig.10). The boundary mesh is represented by a graph $(N, E)$ where each node of $N$ corresponds to a triangle, and each edge of $E$ connects two nodes of $E$ that correspond to triangles that share an edge. Costs are assigned to nodes and edges of $(N, E)$: the cost of assigning a node of $N$, and so a triangle $t$, to each of the 6 labels is proportional to the angle between the normal to $t$ and the label's direction (unary fidelity term); the cost assigned to an edge $e \in E$ is proportional to the angle between the normals of the triangles corresponding to the end points of $e$ (binary term). The optimized solution is adjustable with a compactness over fidelity ratio.

One of our contributions is to propose a tweaked graph-cut labeling that adjusts graph-cut costs on areas that would lead to fragmentation with a naive labeling (Fig.10b). Triangles are in these areas if the two closest labels of their normal have almost the same weight. We use a sensitivity (minimum delta) of $10^{-10}$. For those triangles, before selecting the best label, we apply to the normal a rotation of 0.05 rad around the X, Y and Z axes. Applying the same rotation matrix on the whole mesh give rotation consistency between patches across 90° angles (Fig.10 d). For labeling operators defined in the next section, a labeling of this kind is a better starting point than the original graph-cut labeling. Regarding graph-cut parameters, we use a compactness over fidelity ratio of $\frac{1}{3}$.

## 6 Operators

If the initial labeling is invalid or not all-monotone, we apply modifications to correct it. Modifications are results of a labeling operator applied to solve an invalid component or a turning-point. Some operators were already proposed in the literature (Section 6.1 – Section 6.3), and we introduce new ones (Section 6.4). In this section, we describe operators independently. They will be used all together in an automatic pipeline given in Section 7.

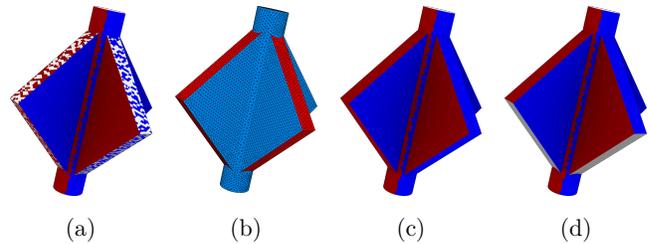

(a)  (b)  (c)  (d)

Figure 10: (a) naive labeling, resulting in charts fragmentation; (b) areas to tilt, in red, with a sensitivity of 0.001; (c) graph-cut labeling, no pre-processing; (d) tweaked graph-cut labeling where our pre-processing adjusts the costs on areas to tilt. B39 model from *MAMBO* [17].

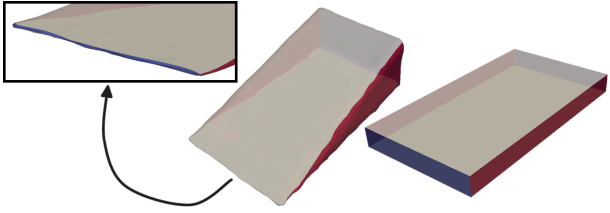

Figure 11: A new chart (top left, in blue) is inserted along a non-orthogonal boundary (middle, between two white charts) to get a valid polycube (right). Illustration from [13, fig.7].

## 6.1 Proposed by Gregson *et al.* 2011 [13]

To our knowledge, [13] were the first to generate a polycube using surface labeling and labeling operators. In their process, after successive volumetric deformations of the input tet mesh, axes alignment is almost achieved. Before the final deformation imposes strict planarity constraints, a naive labeling is computed then edited to obtain a valid polycube topology.

The first modification re-draws jagged boundaries by relabeling adjacent triangles having two times the same label in their neighborhood. Then they target charts having one or two neighbors, and remove them by re-labeling inner triangles starting from the contour (flood-filling). Afterwards, if the labeling contains non-orthogonal boundaries, which by definition forms same-axis adjacent charts, they split these charts with a separating chart along the boundary (Fig.11). The width of the new chart is user-specified, and its label is derived from the labels of its four neighbors, possibly by taking normal similarity into account if there is not enough information to choose a single label.

The labeling analysis also identifies turning-points, which are associated to a wedge at their base (Fig.12 left). At their tip is a highly non-planar and convex chart, which would result in severe distortion under polycube mapping. Axis-aligned cut options are traced on this chart, starting from the turning point, and then evaluated. Valid cuts are those leading to charts with 4 or more neighbors (Fig.12 middle). Among two valid cuts, they select the shortest and most aligned to the global axes. Finally, a chart is emplaced on the cut path (Fig.12), and its label is chosen like the previous operator. Its width is proportional to the width of the base wedge.

## 6.2 Proposed by *PolyCut* [27]

*PolyCut* solutions have a more constraining representation. Labeling modification is not done through direct label assignment updates, but by modifying graph-cut weights. The only operator used in *PolyCut* relies on the local bias of graph-cut weights in a turning-point neighborhood.

## 6.3 Proposed by *Evocube* [31]

The *Evocube* [31] genetic algorithm distinguishes stochastic operators (mutations) and deterministic ones (repairs): re-pairs guarantee a better solution, whereas mutations do not. It is the genetic framework that discards solutions coming from bad mutation choices. We propose a new operators classification, based on what the operator is trying to fix, instead of its deterministic nature in *Evocube*: we distinguish operators that targets validity — polycube topology — and those targeting quality — low-distortion.

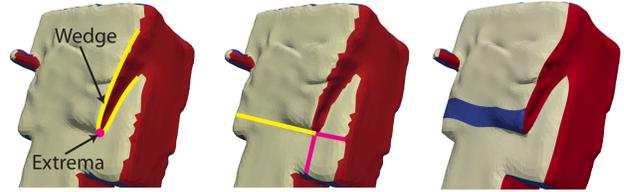

Figure 12: Split of a non-planar chart at a turning-point (left, extrema of a wedge) : among the 3 possible axis-aligned cuts (middle), the yellow one produces a valid labeling, whereas pink ones would produce a chart with only 2 neighbors. Inserting a chart along the yellow cut (right) preserves the red wedge from being collapsed. Illustrations from [13, fig.8].

### 6.3.1 *Evocube* operators targeting the validity

Fig.13 top illustrates these operators within the same order. *Chart removal* is applied on an invalid chart: its facets are fed to a graph-cut optimization, allowing the neighboring labels only, thus forcing the facet re-labeling. *Opposite boundaries* fixes a non-orthogonal boundary by tracing a chart along. *High valence corner* inserts a new chart on what was a corner with a valence greater than 3. Those operators edit the labeling graph to have, locally, a polycube topology. Although not based on a graph-cut optimization, we previously covered how *Gregson et al.* [13] are removing invalid charts in a similar manner. They also process non-orthogonal boundaries like *Evocube* (Section 6.1).

### 6.3.2 *Evocube* operators targeting the quality

Fig.13 bottom illustrates these operators within the same order. Around a turning-point, *chart propagation* spreads the label on one side to the facets of the other side. *Path smoothing* re-draws a boundary to avoid superfluous spikes. These two operators do not modify the labeling graph, only its embedding onto the mesh. *Directional path* creates a chart along a random direction, starting from a turning-point. Note that the polycube topology changes but this operator is still quality-oriented (turning-point removal) and the labeling stays locally valid. The last two are taken from [13] (Section 6.1), but the latter is stochastic within the *Evocube* genetic framework.

## 6.4 New labeling operators

Our analysis of initial labelings (Section 5), taking into account existing operators and the distinction between validity and quality, led us to define new labeling operators.

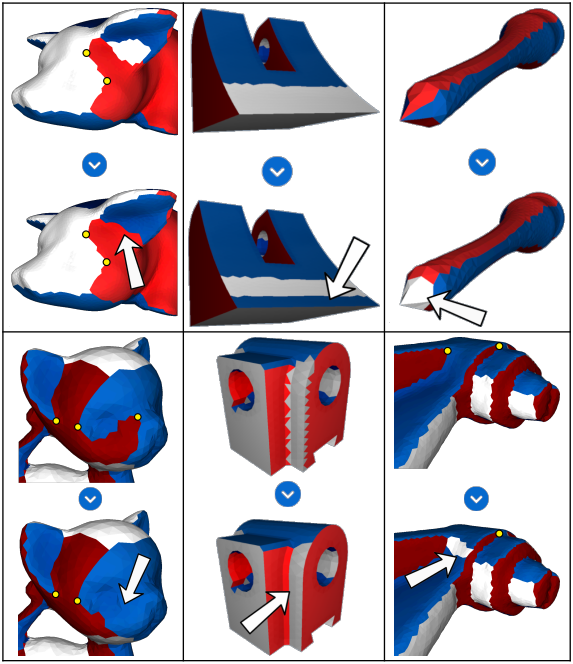

Figure 13: *Evocube* operators (mutations and repairs combined) [31]. From left-to-right then top-to-bottom: *chart removal*, *opposite boundaries*, *high valence corner*, *chart propagation*, *path smoothing* and *directional path*.

#### 6.4.1 Increase chart valence

It is common for CAD models to contain triangular surfaces surrounded by hard edges (Fig.14). The initial labeling is likely to mark the corresponding chart $\mathcal{C}_i$ as invalid (insufficient valence). Such a chart must not be removed, otherwise the quality will be degraded (from both the fidelity and the preservation of feature edges). For such cases, we apply a new operator that identifies the problematic vertex $v$ on an acute angle between same-axis boundary edges, along the contour $C_{\mathcal{C}_i}$ of $\mathcal{C}_i$. Vertex $v$ can be located on a corner (Fig.14a left, in black) or on a turning-point (Fig.14a right, in yellow). With $e_1^v$ and $e_2^v$ the boundary edges incident to $v$ on $C_{\mathcal{C}_i}$, we compute the axis $a \in \{x, y, z\}$ to apply on $C_{\mathcal{C}_i}$, by picking the most aligned to $e_1^v$ and $e_2^v$. This axis will define a new boundary in $C_{\mathcal{C}_i}$ and incrementing the valence of $\mathcal{C}_i$. Two polyedges $p_\circlearrowright$ and $p_\circlearrowleft$ are assembled by aggregating end-to-end boundary edges along $C_{\mathcal{C}_i}$, starting from $v$ and going respectively clockwise and counterclockwise until reaching a corner. Along $p_\circlearrowright$ and $p_\circlearrowleft$, we want to distribute the 2D geometric fidelity between the assignment of the existing label and axis $a$. For each polyedge $p \in \{p_\circlearrowright, p_\circlearrowleft\}$, we compute the cost of assigning every edge to each of the two axes. The association cost is equal to the angle between the edge vector and the label vector. Then, starting from $v$ and going to the other end of $p$, the cumulative cost of $a$ is computed. In the opposite direction, the cumulative cost of the current axis is computed. In this way, an equilibrium point $o_p$ is identified for each $p$. If $v$ does not coincide with one of them, the closer to $v$ is moved on $v$ to enforce the placement of a corner on $v$: between $o_{p_\circlearrowright}$ and $o_{p_\circlearrowleft}$ will be the new boundary around $\mathcal{C}_i$. Afterwards, we greedily trace, edge-by-edge, a boundary $b_o$ starting from each of the two equilibrium point $o$ (Fig.14b). The direction leading this path is along the same axis as the label of $\mathcal{C}_i$. The sign is resolved by checking which one gives the most aligned vector with edges incident to $v$. The path tracing reuses boundary paths if any (Fig.14b left), and stops when a boundary or a turning-point is met. Between the two boundaries $b_o$ we re-label triangles of the base chart $\mathcal{C}_j$, creating a new chart $\mathcal{C}_k$ (Fig.14c). The axis of its label is the remaining one after forbidding the $\mathcal{C}_i$ and the $\mathcal{C}_j$ one. By optimizing the 3D fidelity of the average normal of $\mathcal{C}_k$, we can settle on the label sign.

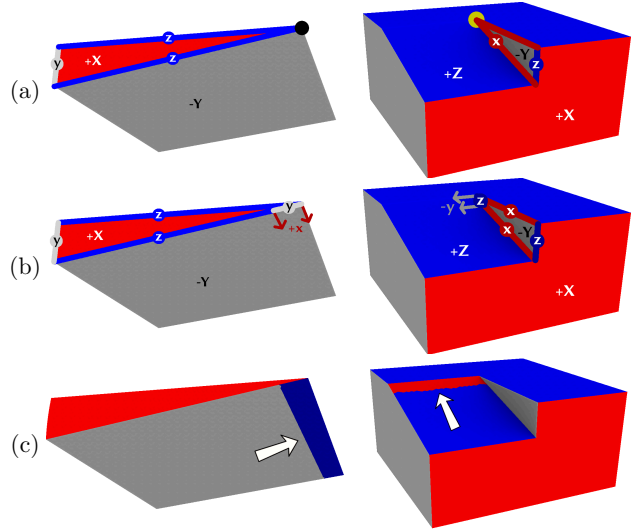

Figure 14: The *increase chart valence* operator applied on *MAMBO* [17] B23 (left) and B49 (right).

#### 6.4.2 Join turning-points pair with new chart

We provide an automatic bond creation between two turning-points $t_1$ and $t_2$, that are at the extremity of lost feature edges. With $T_{\text{left}}$ and $T_{\text{right}}$ the set of facets at the left and right of the feature edges path, the label to insert $l$ is chosen to be different from the 3 labels around $t_1$ and $t_2$, while being close to the average normal of $T_{\text{left}} \cup T_{\text{right}}$. Then we select among $T_{\text{left}}$ and $T_{\text{right}}$ the one having the best fidelity if relabeled with *l*, and we extend the relabeling to adjacent facet of the same side, effectively creating a new chart (Fig.15). Unlike *Evocube* [31] (Section 6.3.2, *directional path*), this operator is deterministic because it is applied on pairs of turning-points.

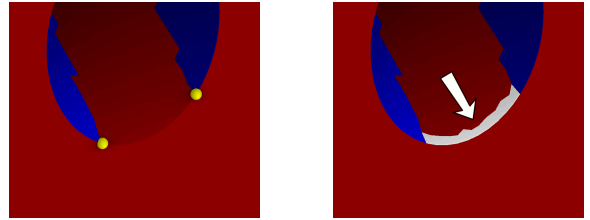

Figure 15: The *join turning-points pair* operator applied on *MAMBO* [17] B29.

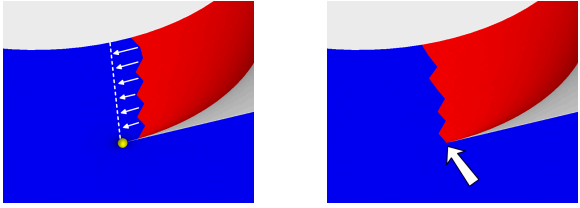

Figure 16: The *pull closest corner* operator applied on *MAMBO* [17] S36. The turning-point is in yellow.

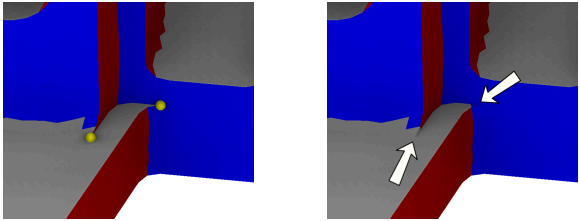

Figure 17: The *move boundary near turning-point* operator applied on CAD5 [45]. Here the two turning-points are removed in one iteration.

#### 6.4.3 Pull closest corner

Our analysis on output labelings shows that some turning-points are on the ideal vertex location for their closest corner *c*. Moving the corner would result in the suppression of the turning-point (Fig.16). The boundary to shift $B_s$ is the next (counter)clockwise incident boundary of *c*, according to the side the turning-point is leaning. A path is greedily traced starting from the turning-point and following the vector $B_s$. Triangles between the current boundary and the new path are relabeled in order to move $B_s$. If feature edges are in the neighborhood of the turning-point, the path is instead traced along the feature edge until we arrive at the corner to move.

#### 6.4.4 Move boundary near turning-point

For turning-points that are not on a feature edge, the process of re-labeling its neighboring triangles moves it closer to the boundary corners. To know which of the two adjacent labels to propagate over the other, we compute the per-label sum of triangle corner angles and keep the maximum (Fig.17).

#### 6.4.5 Straighten boundaries

The labeling graph boundaries $B_i$ being in the end mapped to straight segments in the polycube domain, we have every interest to follow a straight path between the two corners $c_{\text{start}}$ and $c_{\text{end}}$. Thus, we added an operator that re-draw an existing boundary from $c_{\text{start}}$, and step by step choosing the edge leading us closer to $c_{\text{end}}$. Boundaries on feature edges are ignored by this operator and stay on their path.

## 7 Routines

Now, we present how to combine labeling operators in a fully automatic process. In accordance with our goal to favor validity (a polycube topology) over quality (low distortion mapping), we designed two routines (Fig.3): The first one targets invalid charts, boundaries and corners, by inserting and removing charts (Section 7.1); the second one processes non-monotone boundaries (Section 7.2).

### 7.1 Routine to fix the validity

The first operation we apply after the initial labeling (Section 5) is *increase chart valence*, as much as possible. It is done before *fix invalid boundaries* and *fix invalid corners* because

---

**Algorithm 1:** Validity-oriented routine

**Input:** mesh $\partial T_\Omega$, labeling $\ell$, max number of iter. $N_{\max}$
**Output:** updated labeling $\ell$

1 $N \leftarrow 0$
2 **While** $\ell$ is not valid and $N \leq N_{\max}$
3 $\quad N \leftarrow N + 1$
4 $\quad$ **Repeat until** no chart processed
5 $\quad\quad$ Find an invalid chart surrounded by feature edges, if any increase its valence by inserting new chart (Section 6.4.1)
6 $\quad$ **If** $\ell$ is valid **then** return $\ell$
7 $\quad$ **Repeat until** no boundary processed
8 $\quad\quad$ Find the first invalid boundary, if any fix this invalid boundary (Section 6.3.1, *opposite boundary*)
9 $\quad$ **If** $\ell$ is valid **then** return $\ell$
10 $\quad$ **Repeat until** no corner processed
11 $\quad\quad$ Find the first invalid corner, if any fix this invalid corner (Section 6.3.1, *high valence corner*)
12 $\quad$ **If** $\ell$ is valid **then** return $\ell$
13 $\quad$ $S \leftarrow \emptyset$
14 $\quad$ Insert the number of charts, boundaries, corners, invalid chars, invalid boundaries, invalid corners, and turning-point in $S$
15 $\quad$ **While** true
16 $\quad\quad$ Remove all invalid charts (Section 6.3.1, *chart removal*) that are not surrounded by feature edges
17 $\quad\quad$ **If** $\ell$ is valid **then** return $\ell$
18 $\quad\quad$ Compute the number of charts, boundaries, corners, invalid chars, invalid boundaries, invalid corners, and turning-point as *s*
19 $\quad\quad$ **If** $s \in S$ **then** // infinite loop → remove more charts
20 $\quad\quad\quad$ Remove charts around invalid boundaries, like *chart removal* does (Section 6.3.1)
21 $\quad\quad$ **Else** insert *s* in *S*
22 Return $\ell$

configurations on which *increase chart valence* is applied can also contain invalid boundaries and invalid corners, that would be better fixed with *increase chart valence*. We apply the *fix invalid boundaries* then the *fix invalid corners* operator from *Evocube* [31]. The order is important because corners at the extremities of an invalid boundary are likely to be invalid as well, and *fix invalid corners* would not be a proper fix in this case. Finally, we have the most aggressive transformation based on graph-cut [42], [43], [44]: *chart removal* from *Evocube*, on all charts that are not surrounded by feature edges. All of the aforementioned process in embedded in a loop, tracking at each step the number of labeling graph features (charts, boundaries and corners) and the number of invalid ones. If at a given step we encounter for the second time the same set of tracked values, we detect a backtracking and remove the charts around invalid boundaries to escape a potential infinite loop. Between each operator, if we obtain a valid labeling, the routine is stopped (Algorithm 1).

### 7.2 Routine to fix the monotonicity

If we get zero invalid components, we execute a routine to suppress turning-points, aiming to get all-monotone boundaries. We think that getting a valid polycube is more important than getting low parametric distortion, and in this context we delayed the use of the following operators. The one having the most constrained condition of application is the first we apply: *join turning-points pair*. Then, we focus on turning-points near features edges with *pull closest corner*. Then we apply operators for non-monotone boundaries on smooth edges: *move boundary near turning-point* and *straighten boundaries*. The absence of turning-point is checked after applying each operator and allows to stop the routine sooner (Algorithm 2).

## 8 Results

We applied our algorithm on the 113 CAD models of the *MAMBO* dataset [17] and the 109 ones of *OctreeMeshing* [45], after generating tetrahedral meshes with *Gmsh* [18] (characteristic length factor of 0.1) and *MeshGems* [19], respectively, then extracting the surface triangular mesh. A subset of output labelings is shown in Fig.1, and another one is put side by side with *PolyCut* [27] and *Evocube* [31] results in Fig.18 and Table 3. Hexahedral meshes are generated using *HexEx* [21], [46] (scale factor of 1.3), then post-processed with a global padding [2] and a volumetric smoother [47]. Statistical comparisons on the whole datasets are synthesized in Table 1 and Table 2 and per-model results are provided as supplemental material.

For each data subset and each method, we report: the percentage for each four labeling cases; the average labeling fidelity over the whole data subset; the percentage of sharp features edges that are preserved, lost, and ignored (i.e. low dihedral angle)[4]; labeling duration in seconds, including file I/O and including the initial labeling duration for our method, measured on an Intel® Core™ i5-12600H laptop with 32 Go of RAM – through a Windows virtual machine for *PolyCut*; the percentage of hex meshes having a positive minimum Scaled Jacobian; the per-model minimum and average Scaled Jacobian, averaged over the data subset. No hex-meshing was attempted our initial labeling, and for other methods, a hex mesh is generated only if the labeling is valid (all-monotone or not). If despite it all, no hex-mesh is generated, a Scaled Jacobian of $-1$ (worse value) is substituted. *PolyCut* was not executed on *OctreeMeshing* meshes because many of them are beyond the 300k tetrahedra limit of the demo executable. For *MAMBO* we had to generate coarser meshes (characteristic length factor of 0.15).

Our approach stands out for the percentage of valid labeling in all datasets. In particular on *MAMBO/Basic* and *Simple*, our method is the only one to reach 100% of all-monotone and valid labelings. In terms of duration, we offer important speedups (10 to 100 times faster). Regarding quality metrics (Section 3.4), fidelity is similar, and sharp feature edges preservation is a bit worse than our initial labeling, but still better than *Evocube* [31]. On some models, our validity definition (Section 4) significantly improves the labeling and the generation time (Fig.19).

---

**Algorithm 2:** Monotonicity-oriented routine

**Input:** triangle mesh $\partial T_\Omega$, valid labeling $\ell$
**Output:** updated labeling $\ell$

1  **If** all boundaries are monotone **then** return $\ell$
2  **Repeat until** no turning-points processed
3     Find the first pair of turning-points joined by lost feature edges, if any insert a chart between these two turning-points (Section 6.4.2)
4  **If** all boundaries are monotone **then** return $\ell$
5  **Repeat until** no turning-point processed
6     Find the turning-point on feature edges, if any re-trace a boundary of the closest corner, pulling this corner on the turning-point vertex (Section 6.4.3)
7  **If** all boundaries are monotone **then** return $\ell$
8  Process turning-points on smooth surfaces by locally moving their boundary (Section 6.4.4)
9  **If** all boundaries are monotone **then** return $\ell$
10 Straighten boundaries that are not on feature edges (Section 6.4.5)
11 Return $\ell$

---

[4]The *OctreeMeshing* dataset does not contain feature edges, and the *PolyCut* file formats did not allow us to extract such information, hence the empty cells.

| Dataset/subset (size) | Labeling method | Valid and all monotone; Valid with turning-points; Invalid; Failed | | Overall average fidelity | feature edges: Sharp and preserved; Sharp and lost; Ignored | Total labeling duration (s) | minSJ ≥ 0 | Overall average minSJ; avgSJ |
|---|---|---|---|---|---|---|---|---|
| | | ratio | cumulative | | | | | |
| MAMBO/ Basic (74) | Graph-cut | 73.0 %<br>6.8 %<br>20.3 %<br>0.0 % | 73.0 %<br>79.7 %<br>**100.0 %**<br>(100.0 %) | 0.976 | **88.5 %**<br>0.8 %<br>10.6 % | **9.23** | - | - |
| | *PolyCut* [27] | 93.2 %<br>1.4 %<br>0.0 %<br>5.4 % | 93.2 %<br>94.6 %<br>94.6 %<br>(100.0 %) | 0.974 | - | 1 294.15<br>(×78 ours) | 71.4 % | **0.006**<br>0.893 |
| | *Evocube* [31] | 94.6 %<br>1.4 %<br>4.1 %<br>0.0 % | 94.6 %<br>95.9 %<br>**100.0 %**<br>(100.0 %) | 0.966 | 82.3 %<br>7.0 %<br>10.6 % | 2 614.23<br>(×157 ours) | 63.4 % | −0.084<br>**0.925** |
| | Ours | 100.0 %<br>0.0 %<br>0.0 %<br>0.0 % | **100.0 %**<br>**100.0 %**<br>**100.0 %**<br>(100.0 %) | 0.975 | **88.5 %**<br>0.8 %<br>10.6 % | 16.70 | **74.3 %** | −0.042<br>0.922 |
| MAMBO/ Simple (30) | Graph-cut | 40.0 %<br>30.0 %<br>30.0 %<br>0.0 % | 40.0 %<br>70.0 %<br>**100.0 %**<br>(100.0 %) | 0.978 | **81.3 %**<br>1.6 %<br>17.1 % | **4.31** | - | - |
| | *PolyCut* [27] | 86.7 %<br>6.7 %<br>0.0 %<br>6.7 % | 86.7 %<br>93.3 %<br>93.3 %<br>(100.0 %) | 0.977 | - | 842.67<br>(×72 ours) | **57.1 %** | **−0.109**<br>0.907 |
| | *Evocube* [31] | 73.3 %<br>26.7 %<br>0.0 %<br>0.0 % | 73.3 %<br>**100.0 %**<br>**100.0 %**<br>(100.0 %) | 0.974 | 78.8 %<br>4.1 %<br>17.1 % | 1 446.38<br>(×124 ours) | 20.0 % | −0.393<br>**0.925** |
| | Ours | 100.0 %<br>0.0 %<br>0.0 %<br>0.0 % | **100.0 %**<br>**100.0 %**<br>**100.0 %**<br>(100.0 %) | 0.975 | 81.2 %<br>1.7 %<br>17.1 % | 11.67 | 56.7 % | −0.172<br>0.917 |
| MAMBO/ Medium (9) | Graph-cut | 22.2 %<br>33.3 %<br>44.4 %<br>0.0 % | 22.2 %<br>55.6 %<br>**100.0 %**<br>(100.0 %) | 0.966 | **70.5 %**<br>3.6 %<br>25.9 % | **1.26** | - | - |
| | *PolyCut* [27] | 77.8 %<br>22.2 %<br>0.0 %<br>0.0 % | **77.8 %**<br>**100.0 %**<br>**100.0 %**<br>(100.0 %) | 0.963 | - | 409.10<br>(×25 ours) | **22.2 %** | −0.541<br>0.832 |
| | *Evocube* [31] | 22.2 %<br>44.4 %<br>33.3 %<br>0.0 % | 22.2 %<br>66.7 %<br>**100.0 %**<br>(100.0 %) | 0.954 | 65.1 %<br>9.0 %<br>25.9 % | 611.72<br>(×37 ours) | 0.0 % | −0.658<br>0.885 |
| | Ours | 77.8 %<br>11.1 %<br>11.1 %<br>0.0 % | **77.8 %**<br>88.9 %<br>**100.0 %**<br>(100.0 %) | 0.954 | 69.8 %<br>4.3 %<br>25.9 % | 16.51 | 12.5 % | **−0.384**<br>**0.896** |

Table 1: Statistical results on the *MAMBO* dataset.

| Dataset/subset (size) | Labeling method | Valid and all monotone; Valid with turning-points; Invalid; Failed | | Overall average fidelity | Total labeling duration (s) | minSJ $\geq 0$ | Overall average minSJ; avgSJ |
|---|---|---|---|---|---|---|---|
| | | ratio | cumulative | | | | |
| OctreeMeshing/cad (109) | Graph-cut | 70.6 %<br>6.4 %<br>22.9 %<br>0.0 % | 70.6 %<br>77.1 %<br>**100.0 %**<br>(100.0 %) | **0.973** | **8.68** | - | - |
| | *Evocube* [31] | 85.3 %<br>7.3 %<br>7.3 %<br>0.0 % | 85.3 %<br>92.7 %<br>**100.0 %**<br>(100.0 %) | 0.968 | 31 609.29<br>(×570 ours) | 66.3 % | −0.059<br>**0.915** |
| | Ours | 89.0 %<br>8.3 %<br>2.8 %<br>0.0 % | **89.0 %**<br>**97.2 %**<br>**100.0 %**<br>(100.0 %) | 0.963 | 55.49 | **74.5 %** | **0.003**<br>0.913 |

Table 2: Statistical results on the *OctreeMeshing* [45] CAD dataset.

There still are cases where we fail to produce a valid labeling, especially when valid solutions requires to label areas with poor geometric fidelity. On the 113 models from *MAMBO* [17], only the labeling of the helix in Fig.20 is invalid.

As for all fidelity-driven labelings, we are unable to explore more compact polycubes by warping [48]. Despite leading to an higher quality hex-mesh, Fig.21 right is considered a worse solution than Fig.21 left because of high angles between normals and assigned labels. Eventually, by working on the input surface, we cannot detect conflicting normal constraints [36] or in-volume twists [37] that would classify a labeling as valid despite not representing a polycube.

## 9 Conclusion and future work

We presented a novel approach for the labeling stage of automatic polycube generation, based on three main contributions: (1) better formulation of validity criteria (Section 4) that broaden the space of valid polyhedra; (2) biased graph-cut costs to generate initial labelings that better process misaligned parts of the input model (Section 5); (3) a validity-oriented routine and a monotonicity-oriented routine (Section 7) that achieve robustness, while being considerably faster than state-of-the-art algorithms.

We identify future works. The behavior of our boundary tracing algorithm may be improved by taking curvature into consideration and by favoring passing by feature edges with a Dijkstra algorithm. More generally, instead of capturing feature edges when a labeling operator find some, we would benefit from a feature-edge-aware initial labeling. CAD features preservation would be better measured on the dedicated *HexMe* dataset [49]. Because the routines described in Section 7 requires fine-tuning to ensure the conditional branchings to be adapted to a wide range of shapes, a more robust approach may be to use a tree-based exploration algo-rithm, testing different operators orderings, instead of relying on the same routine for all inputs. Finally, our pipeline focuses on labeling validity, then aims at low distortion by considering turning-points. The distortion may be further reduced with new validity-preserving operators, directly driven by the labeling fidelity or by the downstream polycube distortion.

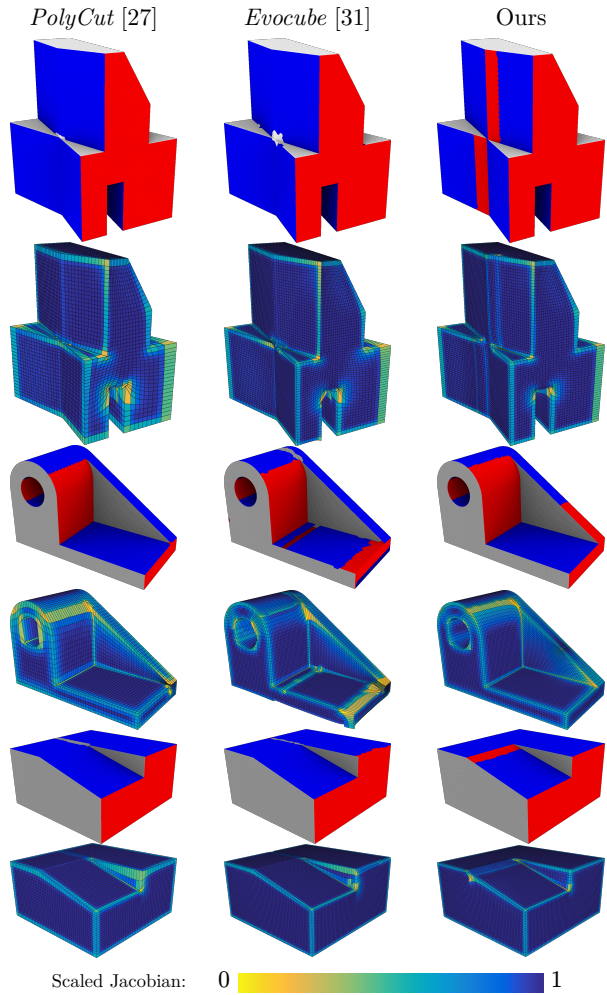

Figure 18: (left to right) Comparisons between *PolyCut* [27], *Evocube* [31] and our algorithm, with hexahedral meshes generated with [21], [46]. Models *MAMBO* [17] B76, S35 and B49.

| model; method | #corners; #invalidities; #turning-points | fidelity: min; avg | preserved feature edges | #hex; minSJ; avgSJ |
|---|---|---|---|---|
| B76 PolyCut | 24 **0** 1 | 0.500 0.980 | – | 29980 −0.879 0.956 |
| B76 Evocube | 26 **0** 0 | 0.500 **0.981** | 91.1 % | 92774 −0.884 0.968 |
| B76 Ours | 34 **0** 0 | **0.711** 0.972 | **95.4 %** | 92877 **−0.025** **0.969** |
| S35 PolyCut | 22 **0** 0 | 0.500 **0.985** | – | 20016 −0.712 0.923 |
| S35 Evocube | 28 **0** 0 | 0.406 0.954 | 79.9 % | 65006 −0.978 0.938 |
| S35 Ours | 22 **0** 0 | **0.749** 0.983 | **99.1 %** | 72974 **0.024** **0.944** |
| B49 PolyCut | 12 **0** 0 | 0.500 0.995 | – | 46110 −0.001 0.989 |
| B49 Evocube | 12 **0** 0 | 0.500 **0.996** | 95.0 % | 147619 −0.022 **0.993** |
| B49 Ours | 14 **0** 0 | **0.629** 0.994 | **100 %** | 151668 **0.069** 0.992 |

Table 3: Statistical results for *PolyCut* [27], *Evocube* [31] and our algorithm, on the three models in Fig.18.

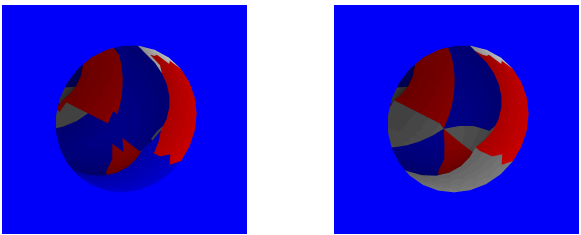

Figure 19: Labelings of *Evocube* [31] (left) and our method (right) on the cheese2 model from [45]. It has 6-connected corners that *Evocube* classifies as invalid and tries to avoid, whereas we rightfully classify them as valid. *Evocube* lasts 4min 36s, while we return a result in 0.127s.

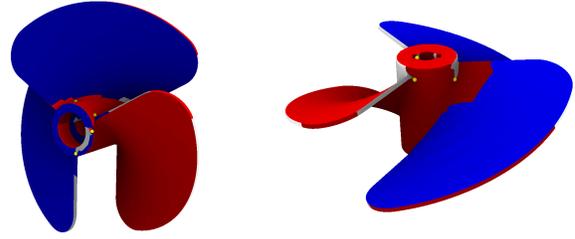

Figure 20: Invalid labeling on *MAMBO* [17] M8.

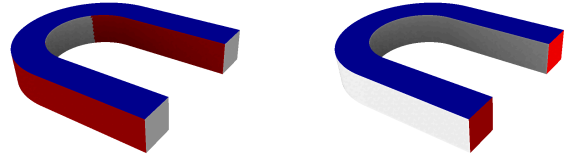

Figure 21: (left) a labeling generated with the fidelity as objective; (right) a labeling corresponding to one cuboid. The latter gives hexahedra that are better surface-aligned.

The polycube distortion can be estimated with fast surface polycube generation [50], [51]. Adding charts and adjusting boundaries placement would noticeably improved downstream hexahedral meshes (Fig.18).

## References

[1] J. F. Shepherd and C. R. Johnson, "Hexahedral mesh generation constraints," *Engineering with Computers*, vol. 24, pp. 195–213, 2008, doi: 10.1007/s00366-008-0091-4.

[2] N. Pietroni *et al.*, "Hex-Mesh Generation and Processing: A Survey," *ACM Trans. Graph.*, vol. 42, no. 2, pp. 1–44, 2022, doi: 10.1145/3554920.

[3] L. Li, P. Zhang, D. Smirnov, S. Mazdak Abulnaga, and J. Solomon, "Interactive all-hex meshing via cuboid decomposition," *ACM Trans. Graph.*, vol. 40, no. 6, 2021, doi: 10.1145/3478513.348056.

[4] F. Zoccheddu, E. Gobbetti, M. Livesu, N. Pietroni, and G. Cherchi, "HexBox: Interactive Box Modeling of Hexahedral Meshes," *Computer Graphics Forum*, vol. 42, no. 5, 2023, doi: 10.1111/cgf.14899.

[5] N. Ray, D. Sokolov, M. Reberol, F. Ledoux, and B. Lévy, "Hex-dominant meshing: Mind the gap!," *Computer-Aided Design*, vol. 102, pp. 94–103, 2018, doi: 10.1016/j.cad.2018.04.012.

[6] R. Schneiders, "A grid-based algorithm for the generation of hexahedral element meshes," *Engineering with Computers*, vol. 5, no. 3, pp. 168–177, 1996, doi: 10.1007/BF01198732.

[7] X. Gao, H. Shen, and D. Panozzo, "Feature Preserving Octree-Based Hexahedral Meshing," *Computer*

*Graphics Forum*, vol. 38, no. 5, pp. 135–149, 2019, doi: 10.1111/cgf.13795.

[8] L. Maréchal, "Advances in Octree-Based All-Hexahedral Mesh Generation: Handling Sharp Features," in *Proceedings of International Meshing Roundtable*, 2009, pp. 65–84. doi: 10.1007/978-3-642-04319-2_5.

[9] A. Sheffer, M. Etzion, A. Rappoport, and M. Bercovier, "Hexahedral Mesh Generation using the Embedded Voronoi Graph," *Engineering with Computers*, vol. 15, pp. 248–262, 1999, doi: 10.1007/s003660050020.

[10] M. Livesu, A. Muntoni, E. Puppo, and R. Scateni, "Skeleton-driven Adaptive Hexahedral Meshing of Tubular Shapes," *Computer Graphics Forum*, 2014, doi: 10.1111/cgf.13021.

[11] P. Viville, P. Kraemer, and D. Bechmann, "Meso-Skeleton Guided Hexahedral Mesh Design," *Computer Graphics Forum*, vol. 42, no. 7, 2023, doi: 10.1111/cgf.14932.

[12] Quadros W. R., "LayTracks3D: A New Approach to Meshing General Solids using Medial Axis Transform," *Procedia Engineering*, 2014, doi: 10.1016/j.proeng.2014.10.374.

[13] J. Gregson, A. Sheffer, and E. Zhang, "All-Hex Mesh Generation via Volumetric PolyCube Deformation," *Computer Graphics Forum*, vol. 30, no. 5, pp. 1407–1416, 2011, doi: 10.1111/j.1467-8659.2011.02015.x.

[14] M. Tarini, K. Hormann, P. Cignoni, and C. Montani, "Polycube-Maps," in *SIGGRAPH*, ACM, 2004, pp. 853–860. doi: 10.1145/1186562.1015810.

[15] M. Nieser, U. Reitebuch, and K. Polthier, "CubeCover - Parameterization of 3D Volumes," *Computer Graphics Forum*, vol. 30, pp. 1397–1406, 2011, doi: 10.1111/j.1467-8659.2011.02014.x.

[16] D. Sokolov, N. Ray, L. Untereiner, and Lévy Bruno, "Hexahedral-Dominant Meshing," *ACM Trans. Graph.*, vol. 35, no. 5, 2016, doi: 10.1145/2930662.

[17] F. Ledoux, *MAMBO: Model dAtabase Mesh BlOcking*. Accessed: May 20, 2024. [Online]. Available: https://gitlab.com/franck.ledoux/mambo

[18] C. Geuzaine and J.-F. Remacle, "Gmsh: a three-dimensional finite element mesh generator with built-in pre- and post-processing facilities," 2008. doi: 10.1002/nme.2579.

[19] Dassault Systèmes - Spatial Corp., "3D Precise Mesh." [Online]. Available: https://www.spatial.com/products/3d-precise-mesh

[20] EDF, CEA, and OpenCascade, "SALOME: The open source platform for numerical simulation." [Online]. Available: https://www.salome-platform.org/?lang=en

[21] M. Lyon, D. Bommes, and L. Kobbelt, "HexEx: Robust Hexahedral Mesh Extraction," *ACM Trans. Graph.*, vol. 35, no. 4, pp. 1–11, 2016, doi: 10.1145/2897824.2925976.

[22] F. Protais, M. Reberol, N. Ray, E. Corman, F. Ledoux, and D. Sokolov, "Robust Quantization for Polycube Maps," *Computer-Aided Design*, vol. 150, 2022, doi: 10.1016/j.cad.2022.103321.

[23] D. Eppstein and E. Mumford, "Steinitz Theorems for Orthogonal Polyhedra," in *Proceedings of the 26th annual symposium on Computational Geometry*, ACM, pp. 429–438. doi: 10.1145/1810959.1811030.

[24] S. Han, J. Xia, and Y. He, "Hexahedral shell mesh construction via volumetric polycube map," *ACM Symposium on Solid and Physical Modeling*, 2010, doi: 10.1145/1839778.1839796.

[25] W. Yu, K. Zhang, S. Wan, and X. Li, "Optimizing polycube domain construction for hexahedral remeshing," *Computer-Aided Design*, vol. 46, pp. 58–68, 2014, doi: 10.1016/j.cad.2013.08.018.

[26] X.-M. Fu, C.-Y. Bai, and Y. Liu, "Efficient Volumetric PolyCube-Map Construction," vol. 35, no. 7, pp. 97–106, 2016, doi: 10.1111/cgf.13007.

[27] M. Livesu, N. Vining, A. Sheffer, J. Gregson, and R. Scateni, "PolyCut: Monotone Graph-Cuts for PolyCube Base-Complex Construction," in *Proceedings of SIGGRAPH Asia*, ACM, 2013. doi: 10.1145/2508363.2508388.

[28] J. Huang, T. Jiang, Z. Shi, Y. Tong, H. Bao, and M. Desbrun, "$\ell$1-Based Construction of Polycube Maps from Complex Shapes," *ACM Trans. Graph.*, vol. 33, no. 3, 2014, doi: 10.1145/2602141.

[29] X. Fang, W. Xu, H. Bao, and J. Huang, "All-Hex Meshing using Closed-Form Induced Polycube," *ACM Trans. Graph.*, vol. 35, no. 4, 2016, doi: 10.1145/2897824.2925957.

[30] Y. Yang, X.-M. Fu, and L. Liu, "Computing Surface PolyCube-Maps by Constrained Voxelization," *Computer Graphics Forum*, vol. 38, no. 7, pp. 299–309, 2019, doi: 10.1111/cgf.13838.

[31] C. Dumery, F. Protais, S. Mestrallet, C. Bourcier, and F. Ledoux, "Evocube: a Genetic Labeling Framework for Polycube-Maps," vol. 41, no. 6, pp. 467–479, 2022, doi: 10.1111/cgf.14649.

[32] L. He, N. Lei, Z. Wang, C. Wang, X. Zheng, and Z. Luo, "Expanding the Solvable Space of Polycube-Map via Validity-Enhanced Construction," in *International Meshing Roundtable*, 2024, pp. 40–52. doi: 10.1137/1.9781611978001.4.

[33] H.-X. Guo, X. Liu, D.-M. Yan, and Y. Liu, "Cut-enhanced PolyCube-maps for feature-aware all-hex meshing," *ACM Trans. Graph.*, vol. 39, no. 4, doi: 10.1145/3386569.3392378.

[34] G. Cherchi, P. Alliez, R. Scateni, M. Lyon, and D. Bommes, "Selective Padding for Polycube-Based Hexahedral Meshing," *Computer Graphics Forum*, vol. 38, no. 1, pp. 580–591, 2019, doi: 10.1111/cgf.13593.

[35] N. Kowalski, F. Ledoux, M. L. Staten, and Owen S. J., "Fun sheet matching: towards automatic block decomposition for hexahedral meshes," *Eng. with Comput*, 2012.

[36] D. Sokolov and N. Ray, "Fixing normal constraints for generation of polycubes," HAL:hal-1211408, 2015. [Online]. Available: https://hal.inria.fr/hal-01211408

[37] S. Mestrallet, F. Protais, C. Bourcier, and F. Ledoux, "Limits and prospects of polycube labelings," 2023, HAL:cea-4169841. [Online]. Available: https://cea.hal.science/cea-04169841

[38] B. Lévy and contributors, *Geogram: a programming library with geometric algorithms*. Accessed: Jun. 24, 2024. [Online]. Available: https://github.com/BrunoLevy/geogram

[39] H. Zhao et al., "Polycube Shape Space," *Computer Graphics Forum*, vol. 38, pp. 311–322, 2019, doi: 10.1111/cgf.13839.

[40] M. Snoep, B. Speckmann, and K. Verbeek, "Polycube Layouts via Iterative Dual Loops," 2024. doi: 10.48550/arXiv.2402.00652.

[41] E. Steinitz, "Polyeder und Raumeinteilungen," *Encyclopädie der mathematischen Wissenschaften*, 1922.

[42] Y. Boykov, O. Veksler, and R. Zabih, "Fast approximate energy minimization via graph cuts," *IEEE Transactions on Pattern Analysis and Machine Intelligence*, pp. 1222–1239, 2001, doi: 10.1109/34.969114.

[43] V. Kolmogorov and R. Zabih, "What Energy Functions can be Minimized via Graph Cuts?," *IEEE Transactions on Pattern Analysis and Machine Intelligence*, pp. 147–159, 2004, doi: 10.1109/TPAMI.2004.1262177.

[44] Y. Boykov and V. Kolmogorov, "An Experimental Comparison of Min-Cut/Max-Flow Algorithms for Energy Minimization in Vision," *IEEE Transactions on Pattern Analysis and Machine Intelligence*, pp. 1124–1137, 2004, doi: 10.1109/TPAMI.2004.60.

[45] "OctreeMeshing dataset." Accessed: Oct. 07, 2024. [Online]. Available: https://cims.nyu.edu/gcl/papers/2019-OctreeMeshing.zip

[46] F. Protais, *polycube with HexEx*. Accessed: May 25, 2024. [Online]. Available: https://github.com/fprotais/polycube_withHexEx

[47] F. Protais, *hexsmoothing*. Accessed: Oct. 04, 2024. [Online]. Available: https://github.com/fprotais/hexsmoothing

[48] M. Mandad, R. Chen, D. Bommes, and M. Campen, "Intrinsic mixed-integer polycubes for hexahedral meshing," *Computer Aided Geometric Design*, vol. 94, 2022, doi: 10.1016/j.cagd.2022.102078.

[49] P.-A. Beaufort, M. Reberol, D. Kalmykov, H. Liu, F. Ledoux, and D. Bommes, "Hex me if you can," *Computer Graphics Forum*, vol. 41, no. 5, pp. 125–134, 2022, doi: 10.1111/cgf.14608.

[50] F. Protais, *fastbndpolycube*. Accessed: Jun. 18, 2024. [Online]. Available: https://github.com/fprotais/fastbndpolycube

[51] H. Zhao, N. Lei, X. Li, P. Zeng, K. Xu, and X. Gu, "Robust edge-preserving surface mesh polycube deformation," *Computational Visual Media*, vol. 4, pp. 33–42, 2018, doi: 10.1007/s41095-017-0100-x.